\documentclass{article}
\usepackage[utf8]{inputenc}
\usepackage[T1]{fontenc}
\DeclareUnicodeCharacter{251C}{\textbar}
\DeclareUnicodeCharacter{2500}{-}
\DeclareUnicodeCharacter{2514}{\textbackslash}
\usepackage{textcomp}
\usepackage{lmodern}
\usepackage{enumitem}
\usepackage{array}
\usepackage{caption}
\usepackage[normalem]{ulem}
\usepackage{biblatex}
\usepackage{graphicx}
\usepackage{xcolor}
\usepackage{fancyhdr}
\usepackage{parskip}
\usepackage{geometry}
\usepackage{changepage}
\usepackage{sectsty}
\allsectionsfont{\sffamily}

\usepackage{todonotes}
\usepackage{verbatim}
\usepackage{listings}

\usepackage{tikz}
\usetikzlibrary{shapes.multipart, positioning, arrows.meta}

\usepackage{authblk}

\makeatletter
\renewcommand\AB@affilsepx{\protect\\}  
\makeatother

\usepackage{titlesec}

\title{Procedural Knowledge Libraries: Towards Executable (Research) Memory}
\author{Hamidah Oderinwale\thanks{McGill University. Contact: \texttt{hamidah.oderinwale@mail.mcgill.ca}}}
\date{June 2025}

\begin{document}
\maketitle
\begin{quote}
``Talk is cheap. Show me the code.'' ---Linus Torvalds
\end{quote}

\begin{abstract}
Procedural Knowledge Libraries (PKLs) are frameworks for capturing the full arc of scientific inquiry, not just its outcomes. Whereas traditional libraries store static end products, PKLs preserve the process that leads to those results, including hypotheses, failures, decisions, and iterations. By addressing the loss of tacit knowledge---typically buried in notebooks, emails, or memory---PKLs lay a foundation for reproducible, collaborative, and adaptive research. PKLs provide executable, version-controlled records that contextualize each step of a research process. For example, a researcher using Jupyter notebooks could share not just final outputs, but also the reasoning, discarded approaches, and intermediate analyses that informed them. This work proposes a framework for implementing PKLs within the Jupyter ecosystem, supported by a lens-based transformation model and procedural storage schema.\footnote{The author thanks David Spivak and Jinglin Li for their feedback on earlier drafts. Large language models (LLMs) were used to assist with writing, conceptual development, and figure development.}
\end{abstract}

\section{Introduction}

Many valuable research artifacts are absent from modern libraries---not because they lack importance, but because current systems struggle to accommodate them. Designing a library is a forward-looking endeavor: it requires anticipating how information will be interpreted, reused, and built upon in the future.

This work is part of a broader agenda: identifying knowledge worth preserving---especially that which supports reuse and transparency---and rethinking how libraries must evolve to capture it. It builds on prior efforts advocating for the preservation of reasoning and experimentation, not just final outcomes \cite{Oderinwale_2025}.

Procedural Knowledge Libraries (PKLs) represent a concrete step toward this vision. PKLs are structured, versioned representations of procedural knowledge. They capture not only what was done but also how and why---organized into modular units that can be queried and reused across domain-specific workflows.

Capturing a \emph{process}---a sequence of intentional actions as they unfold---requires real-time documentation that remains aligned with execution. Because steps evolve, goals shift, and edits are made, the most faithful archives treat the process as a dynamic artifact. Effective documentation, then, must go beyond recording outcomes to include the rationale behind them.

To illustrate the distinction between \emph{process} and \emph{procedure}, and between \emph{information} and \emph{knowledge} \cite{Aamodt1995-zp}, consider the contrast between a lab protocol and a researcher's notebook. A protocol describes fixed, validated steps; the notebook captures informal observations, failed attempts, and evolving reasoning---the researcher’s working understanding of how to adapt those steps. Put simply: a \emph{process} is to a \emph{protocol} as a \emph{procedure} is to a \emph{notebook}.


Similarly, a recipe lists ingredients and instructions, but a chef draws on procedural knowledge to adjust for taste, freshness, and timing under uncertain conditions \cite{Parkinson2012-be}. Or compare a software tutorial with learning an experienced developer’s workflow through pair programming: the tutorial offers a static path, while the workflow reveals how to navigate uncertainty, debug creatively, and optimize using tacit heuristics \cite{Gregory2016-zk}.

These omissions---of intent, trial-and-error, and tacit judgment---stymie collaboration, reproduce redundant effort, and obscure provenance. PKLs address these gaps by enabling earlier-stage collaboration, structured reuse, and clearer methodological accountability. The goal is to construct documentation that captures procedural knowledge: how next steps were inferred, why paths were abandoned, and what informed decisions beyond the trace---where \emph{traces} refer to recorded actions.

\section{Defining Procedural Knowledge Libraries}
In considering the impact of PKLs, a question arises: what is documentation? Especially within the context of Procedural Knowledge Libraries, it's essential to recognize that documentation serves multiple purposes. It can describe the features of an object, provide instructions for its use, or detail its functions. However, in the realm of procedural knowledge, documentation should aim to capture not only the steps involved in a process but also the underlying reasoning, decision points, and adaptations made during its execution.

Procedural knowledge refers to an agent's understanding of how to accomplish specific tasks through structured actions. Unlike conceptual knowledge, which concerns abstract principles and generalizable ideas \cite{McCormick1997-jo}, procedural knowledge consists of executable steps tied to concrete outcomes \cite{byrnes1991role}. I distinguish between procedural information---raw sequences of actions---and true procedural knowledge, which integrates both the steps themselves and their contextual purpose in a domain.

Procedural knowledge is inherently goal-oriented---directions gain meaning only in relation to a specific end goal. This contextual grounding supports domain-specific problem solving but limits transferability, as procedures are often internalized through experience rather than explicitly documented. Capturing this knowledge can help preserve tacit expertise that would otherwise be lost.

A complete understanding of procedural knowledge requires examining both successful and failed processes. While metrics can evaluate a procedure's effectiveness, even unsuccessful attempts contain valuable insights for those who can interpret them. This perspective informs our approach to procedural knowledge libraries, where documenting the relationship between methods and outcomes--including dead ends---building more than just repositories, but libraries that can be used to construct maps of problem-solving instances. 

Procedural knowledge forms the backbone of all goal-directed systems, yet capturing and reusing it remains both challenging and often overlooked.
The core difficulty lies in developing encodings that preserve not only the steps executed but also their contextual rationale---transforming raw execution traces into reproducible, interpretable workflows with their underlying reasoning intact  \cite{Elsaka2017-di}.  While some processes follow clearly articulated hypotheses, others may emerge from intuition or ad-hoc judgment. PKLs cannot directly capture tacit knowledge, but the data they contain and the clarity they bring to a process can make it easier to discern between intentional, iterative decisions and outcomes shaped by intuition, serendipity, or accident.


This challenge is particularly important because, as psychological research demonstrates, procedural learning is inherently more difficult than conceptual learning \cite{Byrnes1991-ug}. The implications are universal: whether in scientific research, technical workflows, or organizational operations, the ability to encode, retrieve, and compare procedural knowledge could unlock new paradigms in how one documents, teaches, and builds upon processes.

To address this, I propose infrastructure that captures procedural knowledge in its full context---the how and why behind methodological choices---transforming isolated expertise into reusable, composable units. 

\section{Contributions}

PKLs serve as containers for domain-specific expertise, designed for transfer and reuse---not just as version trees, but as semantically enriched structures with context-aware annotations on each commit.

This paper makes three contributions: (1) it defines Procedural Knowledge Libraries (PKLs) as a framework for capturing and reusing process-level knowledge in computational workflows; (2) it introduces a lens-based abstraction for selective encoding and transformation of procedural units; and (3) it proposes an architecture for PKLs that includes storage for procedural data, a format for procedural ``diffs'' (patch templates), and a sketch of how procedural data may be queried.

PKLs are presented as a conceptual schema, a working prototype, and a candidate standard for representing procedural knowledge---each contribution reflecting a different level of maturity in the framework’s development.

This paper targets practitioners, tool builders, and infrastructure designers seeking to expand procedural documentation capabilities.






\subsection{Completeness \& Reconstruction}

Completeness in PKLs is relational, not absolute. A PKL artifact is considered complete if a domain-qualified interpreter can reconstruct the procedure in a way that aligns with its original intent. Reconstruction failures may be due to the artifact itself or to gaps in the contextual knowledge of the interpreter \cite{cohen2024datareconstructiondont}.

A \emph{qualified interpreter} can be defined as an agent with domain knowledge comparable to that of the original author. For example, two interpreters running the same plotting script on different datasets will produce different results. This is not because the procedure is incomplete, but because their contexts diverge.

Since the interpretive context is difficult to measure directly, we can instead compare memory-derived procedures with the artifacts generated during execution. Discrepancies between them can reveal missing assumptions or gaps in understanding. Procedural Knowledge Libraries (PKLs)---structured, high-integrity representations of process knowledge---help surface these gaps by linking each step to its semantic and temporal context. This referential structure supports targeted debugging when comprehension breaks down or fidelity to the original process is compromised.

\section{Relevant Work}
Research on episodic memory (EM) offers insight into how experiences are stored and retrieved in the brain. Zeng et al.\ show that one function of EM is using past experiences to prepare for future events \cite{zeng2023modeling}, directly informing our understanding of cognitive replay---an ability that PKLs aim to support. This connects to the concept of \emph{episodic control} \cite{blundell2016modelfreeepisodiccontrol, Ciatto_2025}, where discrete experiences can be retrieved and reused as structured, standalone units.

In distributed systems, Kleppmann’s framework for event classification establishes valuable structural principles \cite{kleppmann2021thinking}. The analysis of event ordering---distinguishing between partial and total order---along with an examination of time-boundedness and persistence characteristics, provides concrete parameters for modeling procedural flows.

In ML, complementary approaches have been developed using Procedural Knowledge Ontologies (PKOs). Carriero et al.\ \cite{carriero2025proceduralknowledgeontologypko} demonstrate how standardized representations make interoperability easier, extending the Linked Terms Methodology for ontology development \cite{Poveda-Villalon2022-hg}. These efforts help create common formats for describing procedural knowledge across different fields.

Further, work by Fan et al. outlines the use of ``chunked sub-routines'' to supplement an agent’s library of ``primitive concepts'' in a process termed ``structured library learning.'' They find that such examples establish a foundation for more efficient learning in novel systems---enabling agents to adapt to new environments more effectively by building on initial library of learned processes. Over time, agents develop ``procedural abstractions,'' referring to increasingly larger fragments of steps, thereby making communication more efficient \cite{mccarthy2021learningcommunicatesharedprocedural}.

Collections of codified processes have also been studied as a means to support compositional learning \cite{defelice2022categoricaltoolsnaturallanguage}, as studied in representation learning, by decomposing workflows into modular, reusable units---allowing models or agents to generalize across tasks by recombining known procedures in novel ways \cite{chang2019automaticallycomposingrepresentationtransformations}.

Furthermore, ML model architectures PRISM (PRocedure Identification with a Segmental Mixture Model) leverage hierarchial Bayesian reasoning to recover procedural abstractions from time-series data in an unsupervised manner \cite{goel2018learning}. By contrast, this work emphasizes determinism and explainability, prioritizing explicit encodings and reproducible transformations over probabilistic inference.

\subsubsection{PKLs for Post Hoc Planning}
PKLs can be understood as a structured analogue to the reverse process of dynamic plan synthesis \cite{abe2025llmmediateddynamicplangeneration, Acharya_2024}---the real-time construction or adjustment of a plan in response to unfolding conditions. In contrast, PKLs focus on generating post hoc representations of procedural logic across a wider range of loosely specified situations. Where dynamic plan reconstruction seeks to infer intent and structure from observed agent behavior, PKLs aim to capture the procedural logic behind complex human tasks. 

Relatedly, ``model diffing'' is an interpretability technique where researchers compare two versions of a model (e.g., a base model and a fine-tuned variant) to identify meaningful differences in behavior or representation \cite{ameisen2025circuit, shah2022modeldiffframeworkcomparinglearning}. What sets interpretability through model diffing and PKLs apart is the stage of model processing they target. Model diffing focuses on changes in weights, features, and internal representations, whereas PKLs operate at a higher level of abstraction, capturing human-readable procedural structures like method, parameter, and design choices.

Despite these differences, the two probing methods are compliments of each other. Model diffing reveals internal changes such as activation shifts or architectural rewiring, while PKL-style diffs capture shifts in intent, strategy, or framing. Both can support \emph{interpretability,} but their units of analysis differ: numerical changes in internal representations versus compositional changes in decision logic. While model diffing traces how a model evolves, PKLs clarify how procedural knowledge develops---and how it can be reused, adapted, or audited. Together, they offer a stronger foundation for model oversight.

\section{PKL System Architecture}

The PKL system centers on three core functions — encoding, storage, and retrieval. 
The diagram below presents a high-level view of how these components interact within the overall architecture.

\begin{figure}[h]
\centering
\begin{tikzpicture}[
  scale=0.85,
  transform shape,
  component/.style={rectangle, draw, minimum height=1cm, minimum width=2.2cm, align=center, font=\small},
  arrow/.style={thick, ->, >=Stealth},
  node distance=1.2cm and 1.6cm
]

\node[component] (encoder) {Encoding Engine\\(ASTs, Diffs, Lenses)};
\node[component, below left=of encoder] (notebooks) {Jupyter Notebooks\\+ Scripts};
\node[component, below right=of encoder] (patchstore) {Patch Store\\(Unified Diffs)};
\node[component, right=1.8cm of encoder] (metadata) {Metadata DB\\(SQLite + Tags)};
\node[component, right=of metadata] (retrieval) {Retrieval Interface\\(CLI/API/UI)};
\node[component, above=of encoder] (lenslib) {Lens Templates\\+ Composition Rules};

\draw[arrow] (notebooks) -- (encoder);
\draw[arrow] (encoder) -- (patchstore);
\draw[arrow] (encoder) -- (metadata);
\draw[arrow] (encoder) -- (lenslib);
\draw[arrow] (lenslib) -- (encoder);
\draw[arrow] (metadata) -- (retrieval);
\draw[arrow] (patchstore) -- (retrieval);

\end{tikzpicture}
\caption{High-level architecture of the PKL system, showing component flow from input to retrieval.}
\label{fig:architecture}
\end{figure}
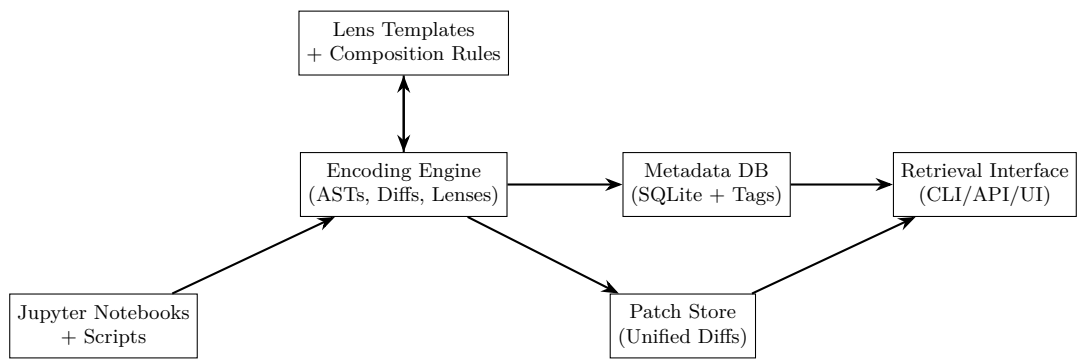

\subsection{Encoding}
\emph{What makes a procedural representation meaningful?}
For PKLs, \emph{encoding} refers to transforming raw procedural traces---such as code cells or pipeline steps---into structured representations that preserve context and support reusability. This section defines what makes an encoding meaningful and how PKLs differentiate between intent and outcome.

\subsection{Definitions}
\begin{itemize}
\item \textbf{Lenses:} Bidirectional transformations that isolate, extract, or reinterpret (procedural) components ``between a database and a view of it.'' \cite{DBLP:journals/corr/abs-1903-03671, Gottlob:CSD-86-258} In PKLs, lenses allow for chosen views, representations, or rewrites of workflows while preserving semantic consistency and enabling composition.
\item \textbf{Procedure:} A versioned, structured representation of a process. In PKLs, a procedure captures the steps, context, and rationale behind a task or workflow, including its evolution over time and potential for reuse or reinterpretation.
\item \textbf{Patches:} In version control systems like Git, a patch is a textual representation of the delta between two states of a repository, typically generated from commits or working tree snapshots. These are commonly expressed as line-based text diffs indicating additions (`+'), removals (`-'), or modifications \cite{joshy2020invariantdiffs}. More broadly, a patch represents a structured set of differences, detailing the transformations required to transition from one state to another. In the context of PKLs, our patch definitions extend this concept to capture not just syntactic deltas but also \emph{semantic differences} and \emph{intended transformations} between procedural units, forming the basis for our lens-based encoding.
\item \textbf{Unit:} The smallest traceable and meaningful element of a procedure. In PKLs, a unit corresponds to an atomic step---such as a code cell, function call, or task---that can be versioned, transformed, or composed within a larger workflow.
\item \textbf{File:} A named unit of data (often a collection of subunits) in a storage system. In PKLs, a file represents a procedural element---such as code, configuration, or output---that forms part of a reproducible workflow.

\item \textbf{Directory:} A container that holds files and/or other directories in a hierarchical structure (for organizational purposes). In PKLs, directories organize procedural artifacts, collections of structured steps or shareable workflows.
\end{itemize}
\subsection{Standardizing Procedural Structure}
Procedures vary across development environments, but the PKL framework is designed to generalize. For any given environment, the system first identifies procedural primitives---the smallest meaningful units of work---such as cells in notebooks, functions in scripts, or tasks in pipelines.

In this system, each unit is stored as a JSON object enriched with metadata. To capture execution structure, abstract syntax trees (ASTs) are used to extract semantic context, while Lamport timestamps record the causal order of operations. This structured representation makes procedures both queryable and interoperable.

The system also defines patch semantics---rules for representing changes such as edits to a cell, reordered steps, or updated parameters. To support targeted reuse and interpretation, it applies lens mappings, which associate user-defined transformations with underlying procedural structures for selective extraction, abstraction, or reinterpretation.

\subsubsection*{Procedural Unit Granularity}
\textit{What is the smallest traceable and meaningful unit of work in the environment?}

\begin{table}[h!]
\centering
\begin{tabular}{|l|l|}
\hline
\textbf{Development Environment} & \textbf{Procedural Unit} \\
\hline
Jupyter Notebooks & Code cell \\
Python Scripts & Function or block \\
RStudio/RMarkdown & Code chunk \\
Data Pipelines (Airflow, Luigi) & Task or DAG node \\
Web IDEs (e.g., Codespaces) & File change or commit \\
Domain-specific Languages & Construct \\
Observable Notebooks & Cell or reactive block \\
\hline
\end{tabular}
\caption{Procedural unit examples across environments}
\end{table}
\subsubsection{Application to Jupyter Workflows}
Encoding a Jupyter notebook into a PKL requires identifying and transforming its procedural elements into structured, queryable units. A notebook consists of code cells, outputs, markdown cells, metadata, and execution order---each of which contributes to the procedural context.

I define the encoding process as follows:

\begin{itemize}
  \item \textbf{Cell Extraction:} Parse the notebook’s JSON to extract code cells, markdown cells, and their metadata.
  \item \textbf{Execution Trace Capture:} Record execution order (e.g., \texttt{execution\_count}) and timestamped outputs to reconstruct the temporal flow.
  \item \textbf{Dependency Analysis:} Infer dependencies between cells using variable usage and definition analysis (e.g., via AST parsing or tools like \texttt{nbdime}).
  \item \textbf{Lens Application:} Apply different lenses such as:
  \begin{itemize}
    \item \textbf{Extraction Lens:} Isolate all cells using a particular library (e.g., \texttt{matplotlib}) or performing a specific task (e.g., data loading).
    \item \textbf{Abstraction Lens:} Collapse multiple exploratory steps into a generalized data-cleaning block.
    \item \textbf{Temporal Lens:} Compare notebook states between two commits or execution runs.
  \end{itemize}
  \item \textbf{Patch Generation:} Represent each edit (cell added, removed, or modified) as a diff in unified format, enriched with semantic tags (e.g., \texttt{type: plotting}, \texttt{intent: exploratory}).
  \item \textbf{Semantic Indexing:} Store the transformed cells and their relationships in the metadata database, enabling retrieval by task, variable, or semantic intent.
\end{itemize}

\textbf{Example Lens Template for Notebook Cell Extraction:}

\begin{verbatim}{{{
"TargetSchema": "cell-snippet-set",
"PatchTemplate": "extractPattern": ".(read_csv | read_excel | read_parquet).",
"cellType": "code",
"matchField": "source"
},
"InversePatchTemplate":
"mergeStrategy": "append-to-top",
"targetNotebook": "inferred"
}
}
\end{verbatim}

\subsection{Towards Meaningful Encodings}

Encoding in a PKL refers to transforming raw procedural information---such as code cells, function calls, or pipeline steps---into structured, reusable representations of procedural knowledge. This section outlines what makes an encoding meaningful and how PKLs differentiate between intent and outcome.

\subsection{Goals \& Challenges}

A central challenge in encoding procedural knowledge is distinguishing between intent and execution. In many workflows---especially exploratory ones, such as computational notebooks---goals often evolve and remain implicit \cite{acm_adam_tabard}. 

PKLs address this by treating the end state---such as a result, figure, or model checkpoint (A saved snapshot of a machine learning model configuration at a specific point during training)---as the source of a linked procedural trace defined as a \emph{procedural unit}.

Each procedural unit (e.g., a notebook cell) may include annotations indicating its purpose or expected outcome (e.g., \texttt{load data}, \texttt{train model}, \texttt{plot results}). These semantic tags, stored in the metadata database, allow us to reconstruct the chain of reasoning and compare what was attempted with what was achieved. For example, a failed training step and its correction can both be stored, annotated with \texttt{intent: model training} and \texttt{outcome: failed convergence}, enabling retrospective understanding.

Encoding the goal explicitly can involve metadata fields like \texttt{TargetMetric}, \texttt{ExpectedOutput}, or \texttt{DesiredState}, while the final artifacts---plots, metrics, exports---can be automatically captured and indexed. This setup makes it possible to query notebooks not only by their outputs (e.g., files) or the code that was written but also by the actions taken and their intended outcomes.

\subsection{Semantic \& Temporal Structuring}

To support meaningful encoding, our PKL system integrates semantic tagging and temporal reasoning through structured metadata. A classification engine parses each unit---such as a Jupyter cell, Python function, or DAG node---using (Python's) abstract syntax trees (ASTs), import resolution, and pattern matching. It labels each unit with semantic tags (e.g., \texttt{data loading}, \texttt{model training}), stored as JSON-LD \cite{jsonldJSONLDJSON}.

\subsubsection{Jupyter Semantic Classification Engine}
Fortunately, many Jupyter notebooks share a common language---Python. Most Python packages are distributed through the Python Package Index (PyPI), a central repository of installable libraries. Each package typically includes documentation, API references, and dependency information. Thus, for a given notebook, the core functions and data processing steps can be extracted by parsing the code cells and identifying function calls, imports, and data operations.

Thus, one can programmatically extract the primitive functions and create a repository of functions to parse from a notebook. Each method can be considered an ``operation'' as a heuristic for a ``step'' or ``unit'' in a PKL. 

The following is a code snippet intended to extract all \texttt{import} statements from a \texttt{.ipynb} file and the function definitions for each.
\begin{verbatim}
import nbformat
import re

# Load the notebook
with open("notebook.ipynb") as f:
    nb = nbformat.read(f, as_version=4)

imports = set()
functions = set()

# Regex patterns for imports and function defs
import_pattern = re.compile(r'^\s*(import|from)\s+[^\s]+')
def_pattern = re.compile(r'^\s*def\s+\w+\s*\$(')

for cell in nb.cells:
    if cell.cell_type == "code":
        for line in cell.source.splitlines():
            if import_pattern.match(line):
                imports.add(line.strip())
            if def_pattern.match(line):
                functions.add(line.strip())

print("Imports found:")
for imp in sorted(imports):
    print(imp)

print("\nFunction definitions found:")
for func in sorted(functions):
    print(func)

\end{verbatim}

Next, module-level abstract syntax trees (ASTs), provided for all Python modules, and accessible through the \texttt{ast} Python module can then be used to produce more granular, expressive annotations for a PKL unit \cite{arts2022towards}. The process consists of parsing Python code in ordered, strcutured fashion producing one hierarchical AST for a whole notebook. ASTs for \texttt{.ipynb} are not inherently cell-order aware. Therefore, the metadata that makes up PKL provenance serves as enrichment to ASTs for procedure extraction.

\section{Order \& Time}
In a notebook, there's different layers of temporality that is valuable to track. First, there's cell-execution order. In a notebook, the order of the cells when editing does not matter. However, someone running a notebook, must execute cells in the appropriate order for it to run correctly. Execution order is already captured in \texttt{.ipynb} file metadata with the \texttt{execution\_count} field which records the order that cells were run in a kernel session. The execution order is represented as an ordered list. 

In PKLs, cell order alone is too coarse-grained, as each cell may contain multiple distinct operations or steps. To address this, the PKL database uses a nested structure, embedding finer-grained PKL units within each notebook cell. Procedural order is tracked using Lamport timestamps---logical clocks that capture causal relationships across operations \cite{Lamport1978}. Each unit is assigned a Lamport value, which is updated with edits, reordering, or forks.

The Lamport timestamp algorithm can be shown as follows: 

\begin{verbatim}
# Initialization
time = 0

# On local event or before sending a message
time = time + 1
send(message, time)

# On receiving a message
(message, timestamp) = receive()
time = max(time, timestamp) + 1
\end{verbatim}

By combining semantic tags with logical timestamps, workflows gain both meaning and a clear sense of order. Logical clocks, such as Lamport timestamps, help track the sequence of events and their causal relationships. Semantic tags add context and make it easier to interpret and organize actions. Together, they allow systems to rebuild the history of a process, handle overlapping edits, and reuse past work in new settings. This makes workflows more reliable, easier to review, and better suited for collaboration.

\subsection{Lenses as Procedural Abstractions}

We propose a patch-based encoding approach in which lenses---bidirectional transformations---are the means for isolating and modifying procedural elements.

\textbf{Lens Types:}

\begin{enumerate}
    \item \textbf{Extraction Lenses:} Isolate specific elements (e.g., hyperparameters)
    \item \textbf{Abstraction Lenses:} Summarize or simplify detailed procedures
    \item \textbf{Transformation Lenses:} Convert formats or representations
    \item \textbf{Temporal Lenses:} Compare or filter by time/version
\end{enumerate}

Each lens is implemented as a parameterized patch template for programmatically constructing sophisticated diffs. These templates preserve reversibility, ensuring that any representational transformation (or view) applied to a procedure can also be undone without loss of information or procedural integrity.
\begin{verbatim}
{
  "LensID": "hyperparameter-focus",
  "Description": "Isolates model hyperparameters from full procedure",
  "SourceSchema": "ml-training-procedure",
  "TargetSchema": "hyperparameter-set",
  "PatchTemplate": {
    "contextLines": 0,
    "pathPattern": "**/model_config.{json,yaml}",
    "extractPattern": "$.training.hyperparameters.*"
  },
  "InversePatchTemplate": {
    "mergeStrategy": "deep-merge",
    "targetPath": "$.training.hyperparameters"
  }
}
\end{verbatim}

\textbf{Lens Composition:}

To support complex operations, lenses can be composed sequentially. Composition is valid when the output schema of one lens matches the input schema of the next.

\begin{verbatim}
def can_compose(lens1, lens2):
    return lens1.targetSchema.isCompatibleWith(lens2.sourceSchema)

def compose_lenses(lens1, lens2):
    if not can_compose(lens1, lens2):
        raise IncompatibleLensError()
    return Lens(
        sourceSchema=lens1.sourceSchema,
        targetSchema=lens2.targetSchema,
        transform=lambda x: lens2.transform(lens1.transform(x)),
        inverse=lambda y: lens1.inverse(lens2.inverse(y))
    )
\end{verbatim}

This modular encoding system allows for scalable transformation and precise interpretation of procedural traces.

\subsection{Storage}
Once encoded, procedural knowledge must be persistently stored in a form that is accessible, queryable, and robust to evolution over time.

\subsubsection{File Format and Storage Considerations}
I leverage established version control concepts and formats:

\begin{itemize}
    \item \textbf{Base Format:} JSON or YAML for structured data, with additional formats for domain-specific data
    \item \textbf{Patch Format:} Standard unified diff format (compatible with git, diff, patch utilities)
    \item \textbf{Directory Structure:}
\begin{verbatim}
procedures/
|- {procedure-id}/
|  |- base/
|  |  \- {files representing base procedure}
|  |- versions/
|  |  |- v1/
|  |  |- v2/
|  |  \- ...
|  \- views/
|     |- {lens-id}-{params-hash}/
|     \- ...
\end{verbatim}
\end{itemize}

\subsubsection{Storage Proposal}
I propose an architecture based on:
\begin{enumerate}
    \item \textbf{Git-Compatible Storage:} Leveraging mature version control for procedure history
    \item \textbf{Patch Files:} Standard unified diff format for representing transformations
    \item \textbf{Metadata Database:} Index for lens information and relationships
\end{enumerate}
\subsection*{Database Schema}
\begin{verbatim}
|- Procedures
|  |- ProcedureID (PK)
|  |- BasePath
|  |- Description
|  |- Tags
|  \- CreatedAt
|- Lenses
|  |- LensID (PK)
|  |- Type (Extraction|Abstraction|Transformation|Temporal)
|  |- SourceSchema
|  |- TargetSchema
|  |- PatchTemplate
|  \- InversePatchTemplate
|- Views
|  |- ViewID (PK)
|  |- ProcedureID (FK)
|  |- LensID (FK)
|  |- Parameters
|  |- Path
|  \- CreatedAt
\- Compositions
   |- CompositionID (PK)
   |- Name
   |- LensSequence [LensID]
   \- ValidationRules
\end{verbatim}
This schema provides a lightweight index on top of a filesystem-based storage system.

\subsubsection{Key Features}

\begin{itemize}
    \item \textbf{Filesystem-Based Storage:} Procedures and their versions are stored as files and directories, making them compatible with existing tools
    \item \textbf{Standard Patch Format:} Changes are represented using the unified diff format, making them human-readable and compatible with existing patch tools
    \item \textbf{Metadata Database:} A database that indexes procedures, lenses, and views for efficient querying
    \item \textbf{Composition Rules:} Explicit rules for lens composition ensure transformations maintain integrity
\end{itemize}

\subsection{Retrieval}
The retrieval layer lets users locate, interpret, and reuse procedural knowledge.

\subsubsection{Query Language}

\begin{verbatim}
# Find procedures by tag
FIND PROCEDURES WHERE tags CONTAINS "image-classification"

# View a procedure through a lens
VIEW procedure-19 THROUGH lens:high_level_summary

# Apply a sequence of lenses
VIEW procedure-19 THROUGH lens:extract_hyperparams THEN lens:visualize_as_table

# Compare versions
DIFF procedure-7:v1 AGAINST procedure-7:v3

# Extract specific steps
FROM procedure-42 GET STEPS 3 TO 5
\end{verbatim}

\subsubsection{Implementing Retrieval}
The retrieval system is implemented as:

\begin{enumerate}
    \item \textbf{Command-Line Interface:} Git-like CLI for managing procedures
    \item \textbf{API Layer:} RESTful and GraphQL APIs for programmatic access
    \item \textbf{Web Interface:} Visual exploration of procedures, versions, and transformations
\end{enumerate}

\subsubsection*{Example CLI commands:}

\begin{verbatim}
# Create a new procedure from existing files
pkl create-procedure --name "image-classification" --base-path ./training_code/

# Apply a lens to create a view
pkl apply-lens hyperparameter-focus --to procedure-42

# Compare two versions
pkl diff procedure-7:v1 procedure-7:v3

# Export a view to a specific format
pkl export procedure-19 --lens high_level_summary --format markdown
\end{verbatim}

\subsection{Practical Implementation}

\subsubsection{File Formats}

\begin{enumerate}
    \item \textbf{Procedure Base:} JSON, YAML, Markdown, code files in native formats
    \item \textbf{Patches:} Standard unified diff format
    \item \textbf{Lens Definitions:} JSON schema
    \item \textbf{Metadata:} SQLite or a fitting alternative
\end{enumerate}

\begin{table}[h!]
\centering
\setlength{\tabcolsep}{10pt} 
\renewcommand{\arraystretch}{1.4} 
\begin{tabular}{|p{3.6cm}|p{3.6cm}|p{3.6cm}|p{3.6cm}|}
\hline
\textbf{Recording a Procedure} & \textbf{Applying a Lens} & \textbf{Composing Lenses} & \textbf{Retrieving and Reusing} \\
\hline
\begin{itemize}[leftmargin=1.2em,itemsep=4pt]
  \item Capture files in native format
  \item Generate unique identifier
  \item Index metadata for searchability
\end{itemize}
&
\begin{itemize}[leftmargin=1.2em,itemsep=4pt]
  \item Select procedure and lens
  \item Generate patch from lens template
  \item Apply patch to create view
  \item Store view and link to source
\end{itemize}
&
\begin{itemize}[leftmargin=1.2em,itemsep=4pt]
  \item Verify lens compatibility
  \item Apply lenses sequentially
  \item Generate composite view
\end{itemize}
&
\begin{itemize}[leftmargin=1.2em,itemsep=4pt]
  \item Query using tags or content
  \item Select views based on context
  \item Export or integrate into workflows
\end{itemize}
\\
\hline
\end{tabular}
\caption{PKL Operation Flow: Four-Stage Process Overview}
\label{tab:process-overview}
\end{table}
\subsubsection{Example Implementation}
\begin{verbatim}
# Creating a procedure
def create_procedure(name, base_path, tags=None):
    # Generate a unique ID
    procedure_id = generate_id()
    
    # Create directory structure
    os.makedirs(f"procedures/{procedure_id}/base")
    os.makedirs(f"procedures/{procedure_id}/versions")
    os.makedirs(f"procedures/{procedure_id}/views")
    
    # Copy base files
    for file in glob.glob(f"{base_path}/**/*", recursive=True):
        if os.path.isfile(file):
            rel_path = os.path.relpath(file, base_path)
            target_path = f"procedures/{procedure_id}/base/{rel_path}"
            os.makedirs(os.path.dirname(target_path), exist_ok=True)
            shutil.copy2(file, target_path)
    
    # Add to index
  db.execute(
    "INSERT INTO Views (ViewID, ProcedureID, LensID, Parameters, "
    "Path, CreatedAt) VALUES (?, ?, ?, ?, ?, ?)",
    (view_id, procedure_id, lens_id, 
     json.dumps(parameters or {}), view_path, datetime.now())
)
    return procedure_id

# Applying a lens
def apply_lens(procedure_id, lens_id, parameters=None):
    # Get procedure and lens info
    procedure = db.query("SELECT * FROM Procedures
    WHERE ProcedureID = ?", (procedure_id,)).fetchone()
    lens = db.query("SELECT * FROM Lenses WHERE LensID = ?", (lens_id,)).fetchone()
    
    # Generate a view ID
    param_hash = hashlib.md5(json.dumps(parameters or {}).encode()).hexdigest()
    view_id = f"{procedure_id}-{lens_id}-{param_hash}"
    view_path = f"procedures/{procedure_id}/views/{lens_id}-{param_hash}"
    
    # Create view directory
    os.makedirs(view_path, exist_ok=True)
    
    # Apply lens template to generate patch
    patch = generate_patch_from_template(
        procedure["BasePath"], 
        json.loads(lens["PatchTemplate"]),
        parameters
    )
    
    # Apply patch to create view
    apply_patch(patch, procedure["BasePath"], view_path)
    
    # Add to index
    db.execute(
        "INSERT INTO Views (ViewID, ProcedureID, 
        LensID, Parameters, Path, CreatedAt)
        VALUES (?, ?, ?, ?, ?, ?)",
        (view_id, procedure_id, lens_id, json.dumps(parameters or {}), view_path, datetime.now())
    )
    
    return view_id
\end{verbatim}

This implementation provides a concrete and practical approach to building a Procedural Knowledge Library system that preserves the essential capabilities for encoding, storing, and retrieving procedural knowledge.
\subsection{Search \& Navigation}

Search is one of the most important features of a Procedural Knowledge Library. It allows users to find and reuse past work without having to manually dig through files or notebooks. Whether someone is looking for a specific step in a project, a general method for solving a problem, or a comparison between different approaches, the search layer makes it possible to retrieve that information quickly and clearly.

Under the hood, search is powered by a structured metadata system. Each piece of a procedure---such as a code cell, function, or data-processing step---is stored as a unit with attached information: what it does, where it came from, when it was created, and how it fits into the larger process. This information is stored in a lightweight SQL database, which support fast indexing and querying.

The core database tables are presented as follows:

\begin{verbatim}
-- Table of procedures
CREATE TABLE Procedures (
    ProcedureID TEXT PRIMARY KEY,
    Name TEXT,
    CreatedAt TIMESTAMP
);

-- Units of procedural work (e.g., cells, functions)
CREATE TABLE Units (
    UnitID TEXT PRIMARY KEY,
    ProcedureID TEXT,
    Content TEXT,
    StartLine INT,
    EndLine INT,
    SemanticTag TEXT,     -- e.g., "data loading", "plotting"
    LamportClock INT,
    FOREIGN KEY (ProcedureID) REFERENCES Procedures(ProcedureID)
);

-- Views created through lens application
CREATE TABLE Views (
    ViewID TEXT PRIMARY KEY,
    ProcedureID TEXT,
    LensID TEXT,
    Parameters TEXT,
    Path TEXT,
    CreatedAt TIMESTAMP
);
\end{verbatim}

Using this database, a range of queries become possible. Here are a few examples:

\begin{itemize}
    \item Search for procedures by task: \texttt{FIND procedures WHERE tag="data cleaning"}
    \item Retrieve steps with a specific role: \texttt{GET all units WHERE role="visualization"}
    \item View a procedure through a different lens: \texttt{VIEW procedure-12 THROUGH lens:abstraction}
    \item Compare versions over time: \texttt{DIFF procedure-19:v1 AGAINST v3}
\end{itemize}

\subsubsection{Search Capabilities}

Search in PKLs is designed not just for finding information, but for helping researchers understand, reuse, and build on prior work.
Each query can return a filtered version of a procedure, a summary of what each part does, or a list of changes between versions. This helps users see not just what was done, but how and why.

To make this accessible, the system includes both a command-line interface (CLI) and a visual interface. The visual interface lets users explore procedures as trees or timelines and apply filters (called “lenses”) interactively. It also supports fuzzy search, so users can type natural phrases like “load CSV and plot scripts” and get relevant results.

\subsubsection{Debugging Example}
Consider a researcher maintaining a model that suddenly underperforms on a key benchmark after recent changes. With a PKL, the researcher can query the last known “good” version and compare it to the current version using a temporal lens:

\begin{verbatim}
DIFF procedure-17:v4 AGAINST procedure-17:v7
\end{verbatim}

This returns not just a code-level diff, but a semantic view---showing, for example, that a hyperparameter tuning step was removed and a data augmentation function was modified. Because each change is tagged with its role (e.g., training config, data preprocessing), the system surfaces meaningful differences rather than low-level line edits.

The researcher can then revisit version 4, review intermediate results, and roll back specific units---streamlining root-cause analysis and avoiding the need to rerun the entire experiment.

\section{The Gaps in Today’s Workflow Tools}

\subsection{Experimental Reproducibility}
Reproducing machine learning results often means guessing how experiments were actually run. Initiatives like the NeurIPS Reproducibility Challenge show the desire for transparency and explainability of the research process. The difficulties faced by the 6-year initiative also show why numerical outputs and shared code are rarely enough \cite{pineau2020improvingreproducibilitymachinelearning}. Tools such as MLflow and the Sacred track parameters, but do not fully capture the steps the researchers took or why \cite{10.1145/3399579.3399867, Greff2017-sx}. Procedural Knowledge Libraries (PKLs) aim to fill this gap by recording not just outputs but also the decisions that led to them.

\subsection{Case Study: Procedural Archaeology in the o1 Model Family}

The OpenAI o1 model family illustrates the need for better process-level documentation. While OpenAI shared results suggesting that inference-time compute---rather than training scale---drove performance gains, no code or method was released to replicate their scaling law graphs \cite{openai2024openaio1card}.

In response, external researchers attempted to recreate OpenAI’s performance on competition math evaluations—without access to the source code \cite{zhang_o1_inference_scaling_laws, openaiLearningReason}. The replicators estimated compute time by prompting the model to think for a specified duration, used billing data to infer token counts, and approximated the majority vote by marginalizing over a sample of reasoning paths \cite{wang2023selfconsistencyimproveschainthought}. Their reproduction came close, but began to diverge at higher compute levels. This gap suggests that some key details—such as evaluation steps or prompt configurations—may not have been shared.

This reconstruction effort---essentially a form of procedural archaeology---shows why PKLs matter. Even when results are public, missing details about how they were produced can lead to confusion and weaken scientific rigor. PKLs help fill this gap by capturing not just the outputs, but also the decisions and steps that led to them.

Evaluating PKLs requires attention not only to theoretical design but also to system overhead. Continuous tracking and versioning introduce encoding, storage, and retrieval costs. While more efficient than duplicating full versions ($O(nv)$), the patch-based approach still incurs retrieval overhead ($O(n \log v)$). Future work should quantify these tradeoffs and explore optimizations in real-world settings.

\section{File Formats \& Interoperability}

PKLs must interoperate with diverse programming environments. Supporting multiple file types is essential if they are to serve as a foundation for rethinking the IDE across languages and workflows. Our system adopts a layered file structure to balance compatibility and expressiveness while minimizing friction for adoption.

At the base layer, native artifacts such as Python scripts, Jupyter notebooks, and configuration files are preserved in their original formats. The diff layer captures changes using the standard unified diff format, enabling version tracking without duplicating entire files. The lens layer encodes semantic transformations using JSON Schema, supporting abstraction and filtering operations across procedures. Finally, the index layer implements a lightweight database schema to enable fast, structured queries over procedural elements.

This structure avoids unnecessary custom formats, enables integration with Git, and supports a wide range of workflows. By wrapping PKL operations as Git extensions, the system enhances familiar tooling with semantic diffs and procedural context---without requiring users to switch environments.

\section{Integration with Existing Systems}

For PKLs to gain adoption, they must integrate smoothly into current research workflows. 

\textbf{Notebooks:} Jupyter is a natural entry point due to its semi-structured format. Our prototype captures execution traces and enables lens-based views via a lightweight extension with minimal workflow disruption.

\textbf{Development Environments:} IDEs emphasize code editing over execution. We address this via language server extensions and inference over execution logs to reconstruct procedural context.

\textbf{Workflow Systems:} Tools like Airflow and Luigi encode execution logic but lack semantic annotations. PKLs augment these systems by layering lens-based views and preserving procedural intent.

\section{Discussion}
\subsection{Adoption Challenges}
PKL adoption faces three barriers: conceptual overhead from the lens-based model, technical overhead from new interfaces, and workflow overhead from integrating procedural capture into existing practices.

A Domain-Specific Language for procedural knowledge creates a trade-off between expressive power and accessibility. Initial implementations should minimize DSL complexity, introducing advanced features as users gain familiarity with core concepts.

\subsubsection{Long-term Viability and Preservation}

The value of procedural knowledge increases over time, making preservation a critical concern. PKLs must be evaluated not just for their immediate utility but for their longevity in the face of changing technologies and research practices.

Our approach to preservation includes: format migration strategies that can adapt as file formats evolve, emulation capabilities that preserve execution environments, semantic annotations that capture intent independent of specific implementations, and dependency management that tracks external resources and their versions.

Thus, PKLs help conserve procedural knowledge so it stays useful---even as tools, formats, and technologies change over time.

\section{Future Work}

A natural direction for future work is to generalize the foundation of Procedural Knowledge Libraries to support a wider range of development environments. Developers frequently move between tools and interfaces---from notebooks and scripts to CI pipelines, configuration files, and visual editors. This raises an open problem: how can PKL artifacts be composed across diverse systems while still enabling procedural capture?

To address this, future PKL systems must support contributions from different tools and environments, and encode transformations that span across modalities and domains. This requires representations that can unify procedures originating from distinct formats and workflows while maintaining structure and interpretability. One promising path lies in research on lenses and bidirectional transformations. Foster et al.\ explore a relevant framework in their work on ``Combinators for Bi-Directional Tree Transformations,'' which addresses the View Update problem in structured data contexts \cite{combinators_trees}. Extending such approaches to PKLs may provide a foundation for composing, diffing, and synchronizing procedural artifacts across tools in a way that remains both interpretable and reversible.

Additionally, a true `collaborative' research notebook ecosystem would support 3-way merging, just as GitHub does. To build on this merge logic (Base, Ours, Theirs), what would it look like to automatically integrate changes that lack structural conflicts, while simultaneously providing an interface for managing more complex conflicts (i.e., two changes that alter the same components or data subsets)? For example, how could graphical representations of an AST subtree or annotated diffs be used to manage or integrate these changes? Furthermore, how would branching work within this paradigm, how would these branches vary in abstraction, and could they be collapsible?

Schemas for extracted procedural data can help ensure consistency in PKL artifacts and help automate the process \cite{distillersr2024data}. Integrating these approaches could automate PKL artifact generation from legacy materials by performing ``procedural'' extraction with standardized metadata schemas \cite{croissant2024metadata}.

Looking ahead, several directions could extend the utility of Procedural Knowledge Libraries (PKLs) and the data they collect. One avenue is their integration with machine learning agents \cite{deepmindAlphaEvolveGeminipowered}, where PKLs could serve as long-term procedural memory---supporting grounded, revisable, multi-step behavior and enabling agents to reflect, adapt, or reuse prior workflows.

\section{Next Steps}
In the context of Jupyter notebooks, PKLs offer a path toward semantically-aware versioning. Traditional line-based diffs are ill-suited to notebooks, which blend code, outputs, markdown, and execution metadata. PKLs can represent these elements as traceable units, supporting structured diffs, procedural patching, and more verifiable forms of citation. Future work includes developing extensions that reconstruct execution graphs from non-linear cell runs \cite{prenner2025simplefaultlocalizationusing}, improving provenance tracking and reproducibility.

Additional work is needed to explore access control and privacy-preserving mechanisms for procedural traces---particularly when they contain sensitive information such as API keys, credentials, or unpublished results. Techniques from differential privacy and secure provenance tracking could be adapted to this setting.

Finally, future efforts should assess the usability of different procedural representations. Comparative studies could examine how formats like notebooks, scripts, or structured workflows affect the ability of users---or models---to understand, modify, or transfer procedures. This research would inform design decisions for PKLs, ensuring they remain not only expressive but learnable and reusable across contexts.

\section{Conclusion}
By capturing both the steps and context of research processes, PKLs allow for more efficient, collaborative, and reproducible science. The architecture I have proposed---based on lens transformations, patch-based storage, and semantic retrieval---provides a practical blueprint for implementing PKLs across diverse research domains. As computational methods continue to dominate scientific discovery, the ability to effectively manage procedural knowledge will become increasingly central to research progress, turning tacit knowledge into explicit, shareable resources that improve how science is done and disseminated. 
\printbibliography

@misc{openaiLearningReason,
  author = {OpenAI},
  title = {{L}earning to reason with {L}{L}{M}s --- openai.com},
  howpublished = {\url{https://openai.com/index/learning-to-reason-with-llms/}},
  year = {2024},
  month = 9,
  day = 12,
  note = {[Accessed 2025-06-16]},
}

@misc{wang2023selfconsistencyimproveschainthought,
      title={Self-Consistency Improves Chain of Thought Reasoning in Language Models}, 
      author={Xuezhi Wang and Jason Wei and Dale Schuurmans and Quoc Le and Ed Chi and Sharan Narang and Aakanksha Chowdhery and Denny Zhou},
      year={2023},
      eprint={2203.11171},
      archivePrefix={arXiv},
      primaryClass={cs.CL},
      url={https://arxiv.org/abs/2203.11171}, 
}

@article{byrnes1991role,
  title={Role of conceptual knowledge in mathematical procedural learning.},
  author={Byrnes, James P and Wasik, Barbara A},
  journal={Developmental psychology},
  volume={27},
  number={5},
  pages={777},
  year={1991},
  publisher={American Psychological Association}
}

@inproceedings{kleppmann2021thinking,
  title={Thinking in events: from databases to distributed collaboration software},
  author={Kleppmann, Martin},
  booktitle={Proceedings of the 15th ACM International Conference on Distributed and Event-based Systems},
  pages={15--24},
  year={2021}
}

@article{zeng2023modeling,
  title={Modeling the function of episodic memory in spatial learning},
  author={Zeng, Xiangshuai and Diekmann, Nicolas and Wiskott, Laurenz and Cheng, Sen},
  journal={Frontiers in Psychology},
  volume={14},
  pages={1160648},
  year={2023},
  publisher={Frontiers Media SA}
}

@misc{carriero2025proceduralknowledgeontologypko,
      title={Procedural Knowledge Ontology (PKO)}, 
      author={Valentina Anita Carriero and Mario Scrocca and Ilaria Baroni and Antonia Azzini and Irene Celino},
      year={2025},
      eprint={2503.20634},
      archivePrefix={arXiv},
      primaryClass={cs.AI},
      url={https://arxiv.org/abs/2503.20634}, 
}

@ARTICLE{Poveda-Villalon2022-hg,
  title     = "{LOT}: An industrial oriented ontology engineering framework",
  author    = "Poveda-Villal{\'o}n, Mar{\'\i}a and Fern{\'a}ndez-Izquierdo,
               Alba and Fern{\'a}ndez-L{\'o}pez, Mariano and
               Garc{\'\i}a-Castro, Ra{\'u}l",
  journal   = "Eng. Appl. Artif. Intell.",
  publisher = "Elsevier BV",
  volume    =  111,
  number    =  104755,
  pages     = "104755",
  month     =  may,
  year      =  2022,
  copyright = "http://creativecommons.org/licenses/by/4.0/",
  language  = "en"
}

@ARTICLE{Byrnes1991-ug,
  title     = "Role of conceptual knowledge in mathematical procedural learning",
  author    = "Byrnes, James P and Wasik, Barbara A",
  journal   = "Dev. Psychol.",
  publisher = "American Psychological Association (APA)",
  volume    =  27,
  number    =  5,
  pages     = "777--786",
  month     =  sep,
  year      =  1991,
  language  = "en"
}

@misc{mccarthy2021learningcommunicatesharedprocedural,
      title={Learning to communicate about shared procedural abstractions}, 
      author={William P. McCarthy and Robert D. Hawkins and Haoliang Wang and Cameron Holdaway and Judith E. Fan},
      year={2021},
      eprint={2107.00077},
      archivePrefix={arXiv},
      primaryClass={cs.CL},
      url={https://arxiv.org/abs/2107.00077}, 
}

@misc{Oderinwale_2025,
  author = {Oderinwale, Hamidah},
  title = {Towards codified context, durable documentation, and process preservation},
  howpublished = {\url{https://topos.institute/blog/2025-02-27-codified-context/}},
  year = {2025},
  month = 2,
  journal = {Topos Institute}
}

@article{DBLP:journals/corr/abs-1903-03671,
  author       = {Brendan Fong and
                  Michael Johnson},
  title        = {Lenses and Learners},
  journal      = {CoRR},
  volume       = {abs/1903.03671},
  year         = {2019},
  url          = {http://arxiv.org/abs/1903.03671},
  eprinttype    = {arXiv},
  eprint       = {1903.03671},
  bibsource    = {dblp computer science bibliography, https://dblp.org}
}

@misc{chang2019automaticallycomposingrepresentationtransformations,
      title={Automatically Composing Representation Transformations as a Means for Generalization}, 
      author={Michael B. Chang and Abhishek Gupta and Sergey Levine and Thomas L. Griffiths},
      year={2019},
      eprint={1807.04640},
      archivePrefix={arXiv},
      primaryClass={cs.LG},
      url={https://arxiv.org/abs/1807.04640}, 
}

@misc{defelice2022categoricaltoolsnaturallanguage,
      title={Categorical Tools for Natural Language Processing}, 
      author={Giovanni de Felice},
      year={2022},
      eprint={2212.06636},
      archivePrefix={arXiv},
      primaryClass={cs.CL},
      url={https://arxiv.org/abs/2212.06636}, 
}

@techreport{Gottlob:CSD-86-258,
  author = {Gottlob, G. and Paolini, P. and Zicari, Roberto},
  title = {Properties and Update Semantics of Consistent Views},
  year = {1986},
  month = 9,
  url = {http://www2.eecs.berkeley.edu/Pubs/TechRpts/1986/6078.html},
  number = {UCB/CSD-86-258},
}

@misc{joshy2020invariantdiffs,
      title={Invariant Diffs}, 
      author={Ashwin Kallingal Joshy and Wei Le},
      year={2020},
      eprint={1911.07988},
      archivePrefix={arXiv},
      primaryClass={cs.SE},
      url={https://arxiv.org/abs/1911.07988}, 
}

@article{Lamport1978,
  title = {Time,  clocks,  and the ordering of events in a distributed system},
  volume = {21},
  ISSN = {1557-7317},
  url = {http://dx.doi.org/10.1145/359545.359563},
  DOI = {10.1145/359545.359563},
  number = {7},
  journal = {Communications of the ACM},
  publisher = {Association for Computing Machinery (ACM)},
  author = {Lamport,  Leslie},
  year = {1978},
  month = jul,
  pages = {558–565}
}

@misc{zhang_o1_inference_scaling_laws,
  author = {Hugh Zhang and Celia Chen},
  title = {Test-Time Compute Scaling Laws},
  year = {2024},
  howpublished = {\url{https://github.com/hughbzhang/o1_inference_scaling_laws}}
}

@misc{pineau2020improvingreproducibilitymachinelearning,
      title={Improving Reproducibility in Machine Learning Research (A Report from the NeurIPS 2019 Reproducibility Program)}, 
      author={Joelle Pineau and Philippe Vincent-Lamarre and Koustuv Sinha and Vincent Larivière and Alina Beygelzimer and Florence d'Alché-Buc and Emily Fox and Hugo Larochelle},
      year={2020},
      eprint={2003.12206},
      archivePrefix={arXiv},
      primaryClass={cs.LG},
      url={https://arxiv.org/abs/2003.12206}, 
}

@inproceedings{10.1145/3399579.3399867,
author = {Chen, Andrew and Chow, Andy and Davidson, Aaron and DCunha, Arjun and Ghodsi, Ali and Hong, Sue Ann and Konwinski, Andy and Mewald, Clemens and Murching, Siddharth and Nykodym, Tomas and Ogilvie, Paul and Parkhe, Mani and Singh, Avesh and Xie, Fen and Zaharia, Matei and Zang, Richard and Zheng, Juntai and Zumar, Corey},
title = {Developments in MLflow: A System to Accelerate the Machine Learning Lifecycle},
year = {2020},
isbn = {9781450380232},
publisher = {Association for Computing Machinery},
address = {New York, NY, USA},
url = {https://doi.org/10.1145/3399579.3399867},
doi = {10.1145/3399579.3399867},
booktitle = {Proceedings of the Fourth International Workshop on Data Management for End-to-End Machine Learning},
articleno = {5},
numpages = {4},
location = {Portland, OR, USA},
series = {DEEM '20}
}

@inproceedings{
goel2018learning,
title={Learning Procedural Abstractions and Evaluating Discrete Latent Temporal Structure},
author={Karan Goel and Emma Brunskill},
booktitle={International Conference on Learning Representations},
year={2019},
url={https://openreview.net/forum?id=ByleB2CcKm},
}

@misc{cohen2024datareconstructiondont,
      title={Data Reconstruction: When You See It and When You Don't}, 
      author={Edith Cohen and Haim Kaplan and Yishay Mansour and Shay Moran and Kobbi Nissim and Uri Stemmer and Eliad Tsfadia},
      year={2024},
      eprint={2405.15753},
      archivePrefix={arXiv},
      primaryClass={cs.CR},
      url={https://arxiv.org/abs/2405.15753}, 
}

@misc{prenner2025simplefaultlocalizationusing,
      title={Simple Fault Localization using Execution Traces}, 
      author={Julian Aron Prenner and Romain Robbes},
      year={2025},
      eprint={2503.04301},
      archivePrefix={arXiv},
      primaryClass={cs.SE},
      url={https://arxiv.org/abs/2503.04301}, 
}

@INPROCEEDINGS{Greff2017-sx,
  title      = "The sacred infrastructure for computational research",
  booktitle  = "Proceedings of the 16th Python in Science Conference",
  author     = "Greff, Klaus and Klein, Aaron and Chovanec, Martin and Hutter,
                Frank and Schmidhuber, J{\"u}rgen",
  publisher  = "SciPy",
  pages      = "49--56",
  year       =  2017,
  conference = "Python in Science Conference",
  location   = "Austin, Texas"
}

@ARTICLE{Gregory2016-zk,
  title     = "Domain model acquisition in domains with action costs",
  author    = "Gregory, Peter and Lindsay, Alan",
  journal   = "Proc. Int. Conf. Autom. Plan. Sched.",
  publisher = "Association for the Advancement of Artificial Intelligence
               (AAAI)",
  volume    =  26,
  pages     = "149--157",
  month     =  mar,
  year      =  2016
}

@ARTICLE{Parkinson2012-be,
  title     = "The application of automated planning to machine tool
               calibration",
  author    = "Parkinson, Simon and Longstaff, Andrew and Crampton, Andrew and
               Gregory, Peter",
  journal   = "Proc. Int. Conf. Autom. Plan. Sched.",
  publisher = "Association for the Advancement of Artificial Intelligence
               (AAAI)",
  volume    =  22,
  pages     = "216--224",
  month     =  may,
  year      =  2012
}

@ARTICLE{Aamodt1995-zp,
  title     = "Different roles and mutual dependencies of data, information,
               and knowledge --- An {AI} perspective on their integration",
  author    = "Aamodt, Agnar and Nyg{\aa}rd, Mads",
  journal   = "Data Knowl. Eng.",
  publisher = "Elsevier BV",
  volume    =  16,
  number    =  3,
  pages     = "191--222",
  month     =  sep,
  year      =  1995,
  language  = "en"
}

@ARTICLE{McCormick1997-jo,
  title     = "Conceptual and procedural knowledge",
  author    = "McCormick, Robert",
  journal   = "Int. J. Technol. Des. Educ.",
  publisher = "Springer Nature",
  volume    =  7,
  number    = "1-2",
  pages     = "141--159",
  month     =  jan,
  year      =  1997
}

@INCOLLECTION{Elsaka2017-di,
  title     = "Fault localization using hybrid static/dynamic analysis",
  booktitle = "Advances in Computers",
  author    = "Elsaka, E",
  publisher = "Elsevier",
  pages     = "79--114",
  series    = "Advances in computers",
  year      =  2017
}

@article{ameisen2025circuit,
  author={Ameisen, Emmanuel and Lindsey, Jack and Pearce, Adam and Gurnee, Wes and Turner, Nicholas L. and Chen, Brian and Citro, Craig and Abrahams, David and Carter, Shan and Hosmer, Basil and Marcus, Jonathan and Sklar, Michael and Templeton, Adly and Bricken, Trenton and McDougall, Callum and Cunningham, Hoagy and Henighan, Thomas and Jermyn, Adam and Jones, Andy and Persic, Andrew and Qi, Zhenyi and Ben Thompson, T. and Zimmerman, Sam and Rivoire, Kelley and Conerly, Thomas and Olah, Chris and Batson, Joshua},
  title={Circuit Tracing: Revealing Computational Graphs in Language Models},
  journal={Transformer Circuits Thread},
  year={2025},
  url={https://transformer-circuits.pub/2025/attribution-graphs/methods.html}
}

@misc{abe2025llmmediateddynamicplangeneration,
      title={LLM-mediated Dynamic Plan Generation with a Multi-Agent Approach}, 
      author={Reo Abe and Akifumi Ito and Kanata Takayasu and Satoshi Kurihara},
      year={2025},
      eprint={2504.01637},
      archivePrefix={arXiv},
      primaryClass={cs.AI},
      url={https://arxiv.org/abs/2504.01637}, 
}

@article{Acharya_2024,
   title={Neurosymbolic Reinforcement Learning and Planning: A Survey},
   volume={5},
   ISSN={2691-4581},
   url={http://dx.doi.org/10.1109/TAI.2023.3311428},
   DOI={10.1109/tai.2023.3311428},
   number={5},
   journal={IEEE Transactions on Artificial Intelligence},
   publisher={Institute of Electrical and Electronics Engineers (IEEE)},
   author={Acharya, Kamal and Raza, Waleed and Dourado, Carlos and Velasquez, Alvaro and Song, Houbing Herbert},
   year={2024},
   month=may, pages={1939–1953} }

@misc{shah2022modeldiffframeworkcomparinglearning,
      title={ModelDiff: A Framework for Comparing Learning Algorithms}, 
      author={Harshay Shah and Sung Min Park and Andrew Ilyas and Aleksander Madry},
      year={2022},
      eprint={2211.12491},
      archivePrefix={arXiv},
      primaryClass={cs.LG},
      url={https://arxiv.org/abs/2211.12491}, 
}

@misc{blundell2016modelfreeepisodiccontrol,
      title={Model-Free Episodic Control}, 
      author={Charles Blundell and Benigno Uria and Alexander Pritzel and Yazhe Li and Avraham Ruderman and Joel Z Leibo and Jack Rae and Daan Wierstra and Demis Hassabis},
      year={2016},
      eprint={1606.04460},
      archivePrefix={arXiv},
      primaryClass={stat.ML},
      url={https://arxiv.org/abs/1606.04460}, 
}

@article{Ciatto_2025,
   title={Large language models as oracles for instantiating ontologies with domain-specific knowledge},
   volume={310},
   ISSN={0950-7051},
   url={http://dx.doi.org/10.1016/j.knosys.2024.112940},
   DOI={10.1016/j.knosys.2024.112940},
   journal={Knowledge-Based Systems},
   publisher={Elsevier BV},
   author={Ciatto, Giovanni and Agiollo, Andrea and Magnini, Matteo and Omicini, Andrea},
   year={2025},
   month=feb, pages={112940} }

@misc{openai2024openaio1card,
      title={OpenAI o1 System Card}, 
      author={OpenAI and : and Aaron Jaech and Adam Kalai and Adam Lerer and Adam Richardson and Ahmed El-Kishky and Aiden Low and Alec Helyar and Aleksander Madry and Alex Beutel and Alex Carney and Alex Iftimie and Alex Karpenko and Alex Tachard Passos and Alexander Neitz and Alexander Prokofiev and Alexander Wei and Allison Tam and Ally Bennett and Ananya Kumar and Andre Saraiva and Andrea Vallone and Andrew Duberstein and Andrew Kondrich and Andrey Mishchenko and Andy Applebaum and Angela Jiang and Ashvin Nair and Barret Zoph and Behrooz Ghorbani and Ben Rossen and Benjamin Sokolowsky and Boaz Barak and Bob McGrew and Borys Minaiev and Botao Hao and Bowen Baker and Brandon Houghton and Brandon McKinzie and Brydon Eastman and Camillo Lugaresi and Cary Bassin and Cary Hudson and Chak Ming Li and Charles de Bourcy and Chelsea Voss and Chen Shen and Chong Zhang and Chris Koch and Chris Orsinger and Christopher Hesse and Claudia Fischer and Clive Chan and Dan Roberts and Daniel Kappler and Daniel Levy and Daniel Selsam and David Dohan and David Farhi and David Mely and David Robinson and Dimitris Tsipras and Doug Li and Dragos Oprica and Eben Freeman and Eddie Zhang and Edmund Wong and Elizabeth Proehl and Enoch Cheung and Eric Mitchell and Eric Wallace and Erik Ritter and Evan Mays and Fan Wang and Felipe Petroski Such and Filippo Raso and Florencia Leoni and Foivos Tsimpourlas and Francis Song and Fred von Lohmann and Freddie Sulit and Geoff Salmon and Giambattista Parascandolo and Gildas Chabot and Grace Zhao and Greg Brockman and Guillaume Leclerc and Hadi Salman and Haiming Bao and Hao Sheng and Hart Andrin and Hessam Bagherinezhad and Hongyu Ren and Hunter Lightman and Hyung Won Chung and Ian Kivlichan and Ian O'Connell and Ian Osband and Ignasi Clavera Gilaberte and Ilge Akkaya and Ilya Kostrikov and Ilya Sutskever and Irina Kofman and Jakub Pachocki and James Lennon and Jason Wei and Jean Harb and Jerry Twore and Jiacheng Feng and Jiahui Yu and Jiayi Weng and Jie Tang and Jieqi Yu and Joaquin Quiñonero Candela and Joe Palermo and Joel Parish and Johannes Heidecke and John Hallman and John Rizzo and Jonathan Gordon and Jonathan Uesato and Jonathan Ward and Joost Huizinga and Julie Wang and Kai Chen and Kai Xiao and Karan Singhal and Karina Nguyen and Karl Cobbe and Katy Shi and Kayla Wood and Kendra Rimbach and Keren Gu-Lemberg and Kevin Liu and Kevin Lu and Kevin Stone and Kevin Yu and Lama Ahmad and Lauren Yang and Leo Liu and Leon Maksin and Leyton Ho and Liam Fedus and Lilian Weng and Linden Li and Lindsay McCallum and Lindsey Held and Lorenz Kuhn and Lukas Kondraciuk and Lukasz Kaiser and Luke Metz and Madelaine Boyd and Maja Trebacz and Manas Joglekar and Mark Chen and Marko Tintor and Mason Meyer and Matt Jones and Matt Kaufer and Max Schwarzer and Meghan Shah and Mehmet Yatbaz and Melody Y. Guan and Mengyuan Xu and Mengyuan Yan and Mia Glaese and Mianna Chen and Michael Lampe and Michael Malek and Michele Wang and Michelle Fradin and Mike McClay and Mikhail Pavlov and Miles Wang and Mingxuan Wang and Mira Murati and Mo Bavarian and Mostafa Rohaninejad and Nat McAleese and Neil Chowdhury and Neil Chowdhury and Nick Ryder and Nikolas Tezak and Noam Brown and Ofir Nachum and Oleg Boiko and Oleg Murk and Olivia Watkins and Patrick Chao and Paul Ashbourne and Pavel Izmailov and Peter Zhokhov and Rachel Dias and Rahul Arora and Randall Lin and Rapha Gontijo Lopes and Raz Gaon and Reah Miyara and Reimar Leike and Renny Hwang and Rhythm Garg and Robin Brown and Roshan James and Rui Shu and Ryan Cheu and Ryan Greene and Saachi Jain and Sam Altman and Sam Toizer and Sam Toyer and Samuel Miserendino and Sandhini Agarwal and Santiago Hernandez and Sasha Baker and Scott McKinney and Scottie Yan and Shengjia Zhao and Shengli Hu and Shibani Santurkar and Shraman Ray Chaudhuri and Shuyuan Zhang and Siyuan Fu and Spencer Papay and Steph Lin and Suchir Balaji and Suvansh Sanjeev and Szymon Sidor and Tal Broda and Aidan Clark and Tao Wang and Taylor Gordon and Ted Sanders and Tejal Patwardhan and Thibault Sottiaux and Thomas Degry and Thomas Dimson and Tianhao Zheng and Timur Garipov and Tom Stasi and Trapit Bansal and Trevor Creech and Troy Peterson and Tyna Eloundou and Valerie Qi and Vineet Kosaraju and Vinnie Monaco and Vitchyr Pong and Vlad Fomenko and Weiyi Zheng and Wenda Zhou and Wes McCabe and Wojciech Zaremba and Yann Dubois and Yinghai Lu and Yining Chen and Young Cha and Yu Bai and Yuchen He and Yuchen Zhang and Yunyun Wang and Zheng Shao and Zhuohan Li},
      year={2024},
      eprint={2412.16720},
      archivePrefix={arXiv},
      primaryClass={cs.AI},
      url={https://arxiv.org/abs/2412.16720}, 
}

@phdthesis{arts2022towards,
  title={Towards Querying Abstract Syntax Trees for Python Programs},
  author={Arts, EM},
  year={2022},
  school={Master Thesis, Department of Computer Science, Eindhoven University of~…}
}

@misc{deepmindAlphaEvolveGeminipowered,
	author = {},
	title = {{A}lpha{E}volve: {A} {G}emini-powered coding agent for designing advanced algorithms --- deepmind.google},
	howpublished = {\url{https://deepmind.google/discover/blog/alphaevolve-a-gemini-powered-coding-agent-for-designing-advanced-algorithms/}},
	year = {},
	note = {[Accessed 22-05-2025]},
}

@inproceedings{acm_adam_tabard,
author = {Rule, Adam and Tabard, Aur\'{e}lien and Hollan, James D.},
title = {Exploration and Explanation in Computational Notebooks},
year = {2018},
isbn = {9781450356206},
publisher = {Association for Computing Machinery},
address = {New York, NY, USA},
url = {https://doi.org/10.1145/3173574.3173606},
doi = {10.1145/3173574.3173606},
booktitle = {Proceedings of the 2018 CHI Conference on Human Factors in Computing Systems},
pages = {1–12},
numpages = {12},
location = {Montreal QC, Canada},
series = {CHI '18}
}

@article{combinators_trees,
author = {Foster, J. Nathan and Greenwald, Michael B. and Moore, Jonathan T. and Pierce, Benjamin C. and Schmitt, Alan},
title = {Combinators for bidirectional tree transformations: A linguistic approach to the view-update problem},
year = {2007},
issue_date = {May 2007},
publisher = {Association for Computing Machinery},
address = {New York, NY, USA},
volume = {29},
number = {3},
issn = {0164-0925},
url = {https://doi.org/10.1145/1232420.1232424},
doi = {10.1145/1232420.1232424},
journal = {ACM Trans. Program. Lang. Syst.},
month = may,
pages = {17–es},
numpages = {65}
}

@misc{jsonldJSONLDJSON,
	author = {},
	title = {{J}{S}{O}{N}-{L}{D} - {J}{S}{O}{N} for {L}inked {D}ata --- json-ld.org},
	howpublished = {\url{https://json-ld.org/}},
	year = {},
	note = {[Accessed 22-05-2025]},
}

@misc{distillersr2024data,
  title={Data Extraction Template for Systematic Review},
  author={{DistillerSR}},
  year={2024},
  howpublished={\url{https://www.distillersr.com/resources/systematic-literature-reviews/data-extraction-template-for-systematic-review}},
  note={Accessed: 2025-05-23}
}

@article{croissant2024metadata,
  title={Croissant: A Metadata Format for ML-Ready Datasets},
  author={Akhtar, Mubashara and Benjelloun, Omar and Conforti, Costanza and others},
  journal={arXiv preprint arXiv:2403.19546},
  year={2024}
}
\end{document}